\documentclass[11pt]{article}
\usepackage[utf8]{inputenc}
\usepackage[margin=1.25in]{geometry}
\usepackage{amsmath, amsfonts, amssymb}
\usepackage{subfig}
\usepackage[dvipsnames]{xcolor}
\definecolor{niceblue}{HTML}{236899}

\usepackage{hyperref}
\usepackage{nameref}
\hypersetup{
    colorlinks=true,
    linkcolor=black,
    filecolor=black,      
    urlcolor=niceblue,
    citecolor=niceblue,
    linkbordercolor = white
}
\usepackage{amsbsy}
\usepackage{epsfig}
\usepackage{subfig}
\usepackage{upgreek}
\usepackage{bm}
\newcommand{\vect}[1]{\bm{#1}}
\newcommand{\mat}[1]{\bm{#1}}
\usepackage{xspace}
\usepackage{color}
\usepackage{subfloat}
\usepackage{lineno}
\usepackage{ragged2e}
\usepackage{placeins}
\usepackage{comment}

\makeatletter
\let\LN@align\align
\let\LN@endalign\endalign
\renewcommand{\align}{\linenomath\LN@align}
\renewcommand{\endalign}{\LN@endalign\endlinenomath}
\let\LN@gather\gather
\let\LN@endgather\endgather
\renewcommand{\gather}{\linenomath\LN@gather}
\renewcommand{\endgather}{\LN@endgather\endlinenomath}
\makeatother

\definecolor{darkgrey}{HTML}{A9A9A9}

\modulolinenumbers[1]

\usepackage{booktabs}
\usepackage[round]{natbib}
\bibpunct{(}{)}{,}{a}{}{,}
\usepackage{authblk}
\usepackage{pdflscape}

\usepackage{textcomp}
\usepackage[sb]{libertine}
\usepackage[varqu,varl]{inconsolata}
\usepackage[libertine,bigdelims,vvarbb]{newtxmath} 
\usepackage[cal=boondoxo]{mathalfa} 
\useosf 
\usepackage[supsfam=libertinesups,supscaled=1.2,raised=-.13em]{superiors}
\usepackage{setspace}

\usepackage{array}
\newcolumntype{^}{>{\currentrowstyle}}

\newcommand{\PreserveBackslash}[1]{\let\temp=\\#1\let\\=\temp}
\newcolumntype{C}[1]{>{\PreserveBackslash\centering}p{#1}}
\newcolumntype{R}[1]{>{\PreserveBackslash\raggedleft}p{#1}}
\newcolumntype{L}[1]{>{\PreserveBackslash\raggedright}p{#1}}

\date{}

\title{
Sequential reduction for discrete latent variables in ecological and evolutionary models using RTMB
}

\author[1,*]{Christopher L. Cahill} 
\author[2,3]{James T. Thorson} 
\author[2]{Kasper Kristensen} 

\affil[1]{Cahill Analytics, Lansing, Michigan, USA}
\affil[2]{Technical University of Denmark, Lyngby, Denmark}
\affil[3]{Alaska Fisheries Science Center, National Marine Fisheries Service, Seattle, USA} %

\begin{document}
\maketitle
\vspace{-1.5em} 
\noindent
\textbf{Corresponding author:} \href{mailto:max.lindmark@slu.se}{christopherfishcahill@gmail.com}

\noindent
\textbf{Acknowledgments:} J. Thorson and C. Cahill are grateful to \citet{royle_hierarchical_2008} for sparking their interest in occupancy and abundance models. The authors thank M. Cheng for a review of this work. 

\noindent
\textbf{Conflict of Interest:} The authors have no conflicts to declare. 

\noindent
\textbf{Author contributions:} K. Kristensen developed the Sequential Reduction algorithm in RTMB, and C. Cahill and K. Kristensen identified its potential role in occupancy and N-mixture models.  J. Thorson lead writing, and developed the phylogenetic trait imputation method.  C. Cahill and J. Thorson developed code for the occupancy and abundance models, and all authors contributed to writing and editing. 

\noindent
\textbf{Data Availability Statement:} All code and data to replicates results are available on GitHub at \url{https://github.com/James-Thorson/SR-RTMB}

\noindent
\textbf{Funding:} The authors have no funding to report.

\noindent
\textbf{Running head:} Sequential reduction for ecologists\\

\clearpage
\setcounter{secnumdepth}{2} 


\section{Abstract}

Statistical models representing ecological and evolutionary dynamics often include latent variables, which can be either continuous- (e.g., average body size) or discrete-valued (e.g., numerical abundance).  Mixed-type hierarchical models (i.e., models with both continuous and discrete random variables) are often fitted using Markov chain Monte Carlo (MCMC), but this can be prohibitively slow for large models.  Here, we introduce an alternative approach in the R package \texttt{RTMB}, which automates the sequential reduction of small groups of related discrete variables, allowing them to be efficiently marginalized for use in hierarchical models. Sequential reduction is then combined with automatic differentiation and the Laplace approximation to efficiently estimate model parameters and generate predictions for both continuous and discrete variables. We demonstrate the speed and flexibility of mixed-type hierarchical models in RTMB using demographic examples (occupancy, dynamic occupancy, N-mixture, open dynamic N-mixture), where we compare RTMB with existing  benchmark software (\texttt{JAGS} and \texttt{unmarked}).  We then develop two novel examples that are feasible using mixed-type hierarchical models.  This first involves a multi-site open N-mixture model with a spatial latent variable for site-specific initial abundance and recruitment.  This example confirms that continuous-valued Gaussian Markov random fields can be estimated jointly with discrete-valued abundance dynamics in less than a minute.  The second involves phylogenetic trait imputation, where we jointly impute a binary variable (viviparity) while estimating its state-switching rates, and also estimate its impact on a continuous variable (body size) during ancestral trait imputation for a published data set of female \textit{Liolaemus} lizards.  This example indicates that phylogenetic comparative methods can estimate linkages among discrete and continuous-valued traits.  We envision that intuitive and efficient specification of mixed-type models will allow more expressive representation of ecological and evolutionary dynamics.

\vspace*{\fill}
\noindent
Keywords: Sequential reduction; hierarchical model; automatic differentiation; Laplace approximation; occupancy model; N-mixture model

\clearpage

\section{Introduction}

Ecologists and evolutionary biologists build bespoke models to represent a wide range of dynamics and address diverse questions. For example, applied ecologists commonly develop purpose-built population dynamics models that not only represent the data availability and hypothesized dynamics of a specific system, but also help inform stakeholders or managers on policy-relevant issues \citep{schaub_lessons_2024}.  Evolutionary ecologists develop phylogenetic models that estimate hypothesized associations among species traits \citep{harmon_phylogenetic_2018}, while community ecologists develop joint species distribution models that leverage demographic and habitat linkages among species \citep{sharma_no_2025}. Despite these disparate aims and applications, each of the problems described above can be formulated as a hierarchical model that links processes of interest and latent variables to observed data.

Over the past twenty years, ecologists have embraced hierarchical modeling, where dynamics are represented as a sequence of conditional distributions among latent (unobserved) variables given estimated rate parameters, and the distribution of data is  specified conditional upon these latent variables \citep{royle_hierarchical_2008}. Estimating hierarchical models requires marginalizing over a distribution of latent variables, which can be achieved using methods such as Markov chain Monte Carlo (MCMC), or approximated using approaches such as the Laplace approximation.  The Laplace approximation is computationally efficient for continuous-valued random effects, where automatic differentiation (AD) can be used to cheaply evaluate the gradient of a user-specified model \citep{skaug_automatic_2006}.  Similarly, AD can be used to improve the sampling efficiency of MCMC, e.g. using Hamiltonian Monte Carlo via Stan \citep{gelman_stan_2015}. However, many models also contain discrete-valued latent variables, which present additional challenges to efficient marginalization. For example, the Laplace approximation is not readily applicable to discrete-valued latent variables, while in Stan, users must often manually derive model-specific methods to integrate out discrete latent states. Discrete-valued variables that evolve along a single dimension (e.g., behavioral state-switching in movement models) can often be handled using hidden Markov models (HMMs), but HMMs become more difficult to apply when multiple discrete-valued states interact or when dependencies occur across multiple dimensions (e.g., spatial interactions represented by an Ising model). 

As one specific example, applied ecologists fit dynamic occupancy models, where occupancy status is estimated as a binary latent variable and samples are treated as Bernoulli draws given this occupancy status \citep{mackenzie_estimating_2002}.  Dynamic (or ''open'') occupancy models then extend this by allowing occupancy status to change over time due to estimated colonization and extirpation rates \citep{mackenzie_estimating_2003}.  Dynamic occupancy models therefore require estimating both discrete-valued occupancy status and continuous-valued rate parameters.  These models are fitted in two main ways:
\begin{itemize}
    \item Bayesian estimation specifies a sequence of conditional distributions for states and data.  Bespoke Bayesian models are easily specified using high-level code such as JAGS \citep{plummer_jags:_2003}, which then implements Gibbs sampling for discrete-valued occupancy status;  
    \item Maximum likelihood (ML) estimation typically requires calculating the probability across states, e.g., using a Hidden Markov model.  This specification requires more user knowledge to develop efficient code, and optimized code is often distributed in packaged form, e.g., \texttt{unmarked} \citep{fiske_unmarked:_2011}. 
\end{itemize}
 Bayesian estimation requires repeated sampling from the posterior distribution and thus is often slower than comparable ML methods. However, high level (i.e., expressive) Bayesian languages provide a flexible framework for specifying mixed-type hierarchical models (i.e., models with both continuous and discrete random variables). In contrast, ML can be faster, but typically requires the development of model-specific algorithms to marginalize over discrete latent states.

In the following, we focus on the Sequential Reduction (SR) algorithm, which enables marginalization over discrete latent states and can be viewed as a generalization of the hidden Markov models (HMMs) already used in ecology.  SR requires identifying a computational tree, where the probability mass function for each discrete variable depends only on a small subset of variables that have been previously estimated \citep{hojsgaard_graphical_2012}.  This ``clique`` of previously computed variables can be automatically detected from the computational tree that is also used during AD.  While some modeling frameworks used by ecologists have adopted automatic differentiation (AD) to improve computational speed and numerical stability, we are not aware of any that also use SR to marginalize over discrete-valued variables within a high-level model-building framework or in combination with integration over continuous-valued latent variables in a mixed-type hierarchical model.

In this paper we introduce a general approach for fitting mixed-type hierarchical models—those containing both continuous and discrete latent variables—via ML using the R package \texttt{RTMB} \citep{kristensen_rtmb_2024}. This approach automates sequential reduction of discrete latent states and combines it with automatic differentiation and the Laplace approximation to enable fast, gradient-based inference as an alternative to MCMC. We first benchmark the approach against existing single-purpose software using four widely used demographic models with discrete latent states (occupancy, dynamic occupancy, N-mixture, and open N-mixture), comparing parameter recovery, runtime, and convergence. We then demonstrate the generality and flexibility of the approach with two bespoke mixed-type models: an open N-mixture model with latent spatial variation in recruitment and initial abundance, and a phylogenetic trait-imputation model that jointly estimates switching rates for a discrete trait (viviparity) and its effect on a continuous trait (body size) along a dated phylogeny. Both models fit in seconds, illustrating the potential for substantially faster inference than analogous MCMC implementations for problems of this size. Across these applications, we demonstrate how automated sequential reduction of discrete latent variables can extend automatic differentiation and RTMB to a broad class of ecological and evolutionary questions that were previously difficult to address, enabling intuitive and efficient specification of mixed-type models and more expressive representations of ecological and evolutionary dynamics.

\section{Methods}

\subsection{Overview}
We seek to develop a computationally efficient process for ecologists to specify a statistical model that includes both continuous- and discrete-valued variables, while allowing them to marginalize across discrete values (via summation) as well as continuous values (via integration), while automatically detecting an efficient ordering for marginalizing variables.  In the implementation described here, the user starts by specifying a function that involves a sequence of computations that result in a joint negative log-likelihood, which is composed of the sum of terms representing the log-probability of discrete- and continuous-valued random effects, and other terms representing the conditional log-likelihood of data.  The user also identifies permissible values for discrete-valued random effects, and RTMB then automatically implements sequential reduction to efficiently sum across all possible combinations.  Afer this, RTMB uses the resulting value to apply the Laplace approximation for the marginal log-likelihood, calculates its gradient with respect to fixed effects, and these can be optimized by gradient-based nonlinear minimizers \citep{kristensen_tmb:_2014,kristensen_rtmb_2024}.  The techniques for the Laplace approximation have been discussed and demonstrated elsewhere, so we here focus upon the sequential reduction algorithm and how it enables efficient estimation for discrete-valued or mixed-type hierarchical models. 

\subsection{Sequential reduction}

Sequential reduction (SR) is a technique to sum a multivariate function $f$ across a high dimensional product space (see Fig. \ref{fig:sequential_reduction}).  In our specific case, $f(\vect{x})$ is the joint likelihood that depends upon an $N$-length vector of discrete-valued random effects $x_n \in G$, where $G$ is the set of permitted values (e.g., $G = \{0,1\}$ for a binary variable representing unoccupied or occupied habitat).  SR requires that $f$ factorizes as a product of a sequence of functions (each termed a ''factor") $f_1(x_{I_1})$, $f_2(x_{I_2})$, ..., $f_k(x_{I_k})$, such that $f$ is ''separable".  Each separate factor then depends upon a smaller subset of variables, i.e., where $f_1(x_{I_1})$ depends upon subset $I_1$ of $\vect{x}$.  We then wish to sum across all possible combinations of discrete-valued variables:

\[\sum_{x_1 \in G,...,x_n \in G} f_1(x_{I_1}) f_2(x_{I_2}) ... f_k(x_{I_k})\]

\noindent across a multivariate grid $G^n$ where each factor, e.g. $f_1(x_{I_1})$ depends on a small number of $x_j$ where $j \in I_1$ and $I_j \subset \{1,...,n\}$.

Given this factorized sequence, SR sums out variables sequentially using a low-dimensional grid of permissible values, while in each step storing the minimal information necessary to carry the summation forward. It can therefore be thought of as the natural generalization of Hidden Markov models (HMMs) to arbitrary graphs. For simplicity, assume we sum out variables in the order $1,...n$. Starting with $x_1$, we locate all factors that depend on $x_1$, or equivalently, all $I_i$ containing the index $1$. The union $U_1 = \cup_i \{I_i : 1 \in I_i\}$ now defines an index set of variables linking to $1$ and we tabulate the corresponding product of factors across the space $G^{|U_1|}$ and sum out $x_1$ (i.e., a discrete convolution across $U_1$). The result is an array indexed by the set $U_1 \setminus \{ 1 \}$ (referred to as a 'clique') representing the integrated function as a new function of $x_{U_1 \setminus \{1\}}$ evaluated on a grid. This is a sufficient representation because integration with respect to future variables are across the same grid. We can therefore proceed recursively by marginalizing across subsequent variables $x_2, ..., x_n$ in order.

The process for marginalizing across (and eliminating) variables has a useful graph interpretation (Fig. \ref{fig:sequential_reduction}). We can define a graph where $x_1,...,x_n$ are vertices, and two variables are connected if they appear in the same factor. The elimination (integration) of a variable, say $x_1$, then removes $x_1$ from the graph, but adds connections between all the neighbors of $x_1$. The number of connections, and thereby array dimensions, can quickly increase beyond practical levels. This phenomenon is the equivalent of 'fill-in' in the sparse Cholesky factorization methods \citep{davis_direct_2006} where a well known remedy is to reorder the graph, i.e. eliminate variables in a different order to reduce the size of the maximum clique. The computational complexity of sequential reduction is $O(n|G|^m)$, where $n$ is the number of random effects, $|G|$ is the number of grid points per random effect, and $m$ is the maximum clique size induced by the elimination order. Thus, for fixed $m$ and $|G|$, the computational cost is linear in the number of random effects and exponential in the maximum clique size. More generally, the cost is polynomial in $n$ only when the maximum clique size does not grow with $n$ (i.e., in a linear chain such as an HMM). In the following, we explore existing and novel mixed-type models for ecology and evolution, and demonstrate that "fill-in'' does not preclude efficient computation for model sizes and complexities arising in many contexts.

\subsection{Performance comparison against existing methods} \label{sec:performance_comparison}

To demonstrate SR, we consider four widely used ecological models that vary along two general dimensions: the representation of the latent state and its temporal structure. The latent state can be either binary or count-valued, while the population can be modeled as either closed over the sampling period or open to demographic changes between sampling intervals. We specifically explore:

\begin{enumerate}
    \item Occupancy models \citep{mackenzie_estimating_2002}, with a binary state-variable $z_j \sim \mathrm{Bernoulli}(\psi)$ representing unoccupied ($z_j=0$) or occupied ($z_j=1$) status for each site $j \in \left( 1, 2, ..., J \right)$.  Occupancy probability $\psi$ is then estimated using on replicated samples $y_{ij} \sim \mathrm{Bernoulli}(pz_j)$ that follow a Bernoulli distribution given the latent state and estimated detection probability $p$.  In the following we explore $J=200$ sites with $I=5$ replicates, with occupancy rate $\psi=0.2$ and detection probability $p=0.5$;

    \item Dynamic occupancy models \citep{mackenzie_estimating_2003}, with a binary state-variable $z_{jt} \sim \mathrm{Bernoulli}(\psi)$ representing
    the unoccupied ($z_{jt}=0$) or occupied ($z_{jt}=1$) status of site
    $j \in \left(1, 2, ..., J\right)$ in season $t = 1$, which then
    transitions as $z_{j,t+1} \sim \mathrm{Bernoulli}(z_{jt}\omega +
    (1-z_{jt})\lambda)$ given persistence probability $\omega$ and
    colonization probability $\lambda$.  Occupancy dynamics are then
    estimated using replicated samples $y_{jti} \sim
    \mathrm{Bernoulli}(pz_{jt})$ given detection probability $p$.  In
    the following we explore $J=200$ sites across $T=4$ seasons with
    $I=4$ replicates, with initial occupancy $\psi=0.6$, persistence
    $\omega=0.8$, colonization $\lambda=0.2$, and detection probability
    $p=0.5$;
    
    \item N-mixture models \citep{royle_n-mixture_2004}, with a categorical state-variable representing numerical abundance $z_j \sim \mathrm{Poisson}(\lambda)$.  Estimated density $\lambda$ is then estimated using replicated samples $y_{ij} \sim \mathrm{Binomial}(z_j, p)$ given detection probability $p$.  In the following we explore $J=100$ sites with $I=5$ samples per site, with expected density $\lambda=32$ and detection probability $p=0.25$;

    \item Open N-mixture models \citep{dail_models_2011}, with a state-variable $z_{jt}$ representing numerical dynamics for each site $j \in \left(1, 2, ..., J\right)$ and time $t\in \left( 1,...,T \right)$. This model specifies a distribution for initial abundance $z_{j1} \sim \mathrm{Poisson}(\lambda)$, where abundance dynamics arise from survival $s_{jt} \sim \mathrm{Binomial}(z_{jt}, \omega)$ and recruitment $g_{jt} \sim \mathrm{Poisson}(\gamma)$, where $z_{j,t+1} = s_{jt} + g_{jt}$. Initial abundance $\lambda$, recruitment rate $\gamma$, and survival rate $\omega$ are then estimated using counts $y_{jt} \sim \mathrm{Binomial}(z_{jt}, p)$ given detection probability $p$. In the following we explore $J=100$ sites over $T=5$ times, with initial abundance $\lambda=4$, recruitment $\gamma=1.5$, and survival $\omega=0.8$, with individual detection probability $p=0.5$. For this model, we used a joint parameterization of the discrete latent variables which is described in S1. 
\end{enumerate}

\noindent We note that marginalizing across states is easy to implement for the occupancy and N-mixture models, as it requires a summation across occupancy status or abundance for each site individually (i.e., SR involves a one-dimensional clique).  However, marginalizing across states in the open N-mixture model involves a discrete convolution of $z_{j,t+1}$, $z_{jt}$, and $g_{jt}$, i.e., the minimum clique for SR is three variables (and we marginalize abundance variables across $G \in \left( 0,1,...,K \right)$ where upper bound $K$ is three times the maximum observed count across sites and times).  In most bespoke models, the three-dimensional convolution would require careful implementation, but this is automatically detected by RTMB. 
    
We specifically compare SR performance as implemented in RTMB against existing maximum likelihood (using \texttt{unmarked}) and Bayesian implementations (using \texttt{JAGS}). For each model and simulation replicate ($n = 100$) we record the maximum likelihood or posterior mean estimate for rate parameters to compare with known true values, and also record the time to convergence for all three estimation methods (including standard error estimates for RTMB). For maximum likelihood models, we consider a model to have converged when the optimizer reports successful convergence (i.e., an \texttt{nlminb} exit code of 0 for RTMB or the corresponding successful convergence code for \texttt{unmarked}) and the resulting Hessian matrix is invertible. For Bayesian models, we assess convergence using the Gelman--Rubin diagnostic, requiring all $\hat{R}$ statistics to be less than 1.1. However, for the N-mixture and open N-mixture models, which exhibited long run times and poor convergence with uninformative priors, we retained JAGS estimates regardless of whether this criterion was met. Applying the $\hat{R} < 1.1$ criterion to these models would have resulted in too few replicates for meaningful comparison. For these specific models, a simulation replicate was therefore retained provided that the corresponding RTMB and \texttt{unmarked} models successfully converged and their Hessian matrices were invertible.

\subsection{Spatial open N-mixture model}

To demonstrate the potential value of mixed-type hierarchical models estimated via RTMB, we develop the first (to our knowledge) model that includes open abundance dynamics across multiple sites, where site-specific recruitment (and resulting abundance) varies among sites following a continuous-valued spatial latent variable. We specifically extend the open N-mixture model to include a spatial latent variable $\vect{\epsilon} \sim \mathrm{GMRF}(\vect{0},\mat{Q})$, where $\epsilon_s$ represents the latent spatial effect at mesh vertex $s$, and $\vect{\epsilon}$ follows a Gaussian Markov random field with sparse precision $\mat{Q}$ constructed using the Stochastic Partial Differential Equation (SPDE) method \citep{lindgren_explicit_2011}. Site-level effects are obtained by projection, $\epsilon_j = \sum_s A_{js}\,\epsilon_s$, where $\mat{A}$ maps the mesh to observed site locations. The SPDE parameterization defines $\kappa$ as the parameter controlling the spatial range of the Matérn correlation and $\tau$ as the parameter controlling the magnitude of spatial variation. We specify site-specific variation in habitat quality, which causes variation in expected initial abundance $\lambda_j = \mu_{\lambda}\times e^{\epsilon_j}$ and subsequent recruitment $\gamma_j = \mu_{\gamma} \times e^{\epsilon_j}$. We use SR to marginalize over the discrete abundance states $z_{jt}$, followed by the Laplace approximation to marginalize over the continuous spatial effects $\vect{\epsilon}$. In other words, SR performs the summation over the possible abundance states, while the Laplace approximation provides an approximation to the integral over the continuous spatial field. RTMB then automatically detects and exploits sparsity in the Hessian of the joint negative log-likelihood with respect to $\vect{\epsilon}$, inherited from the sparse structure of $\mat{Q}$, so that computational cost scales with the number of nonzero entries rather than with the square of the number of sites \citep{lindgren_explicit_2011}. Thus, models combining discrete abundance states and continuous spatial effects can be fitted by marginal maximum likelihood without manually deriving marginalization procedures or implementing a custom sampling algorithm.

To evaluate performance, we conduct a simulation experiment involving $T=4$ times and $J=200$ sites that are randomly distributed with a square $1\times1$ domain. We then use the \texttt{fmesher} package \citep{lindgren_fmesher_2023} to construct matrices $\left( \mat{M}_0, \mat{M}_1, \mat{M}_2 \right)$ that are used to estimate the spatial precision matrix $\mat{Q} = \tau^2 \left( \kappa^4 \mat{M}_0 + 2\kappa^2 \mat{M}_1 + \mat{M}_2 \right)$, where we simulate $\kappa = 9.4$ and $\tau = 0.06$ (corresponding to a decorrelation distance of $0.3$ and pointwise standard deviation of $0.5$). We then simulate expected initial abundance $\lambda=2.5$, recruitment $\gamma=1.5$, and survival $\omega=0.7$, with individual detection probability $p=0.4$. For each simulation replicate, we record the estimated parameters to compare with known true values, and record the optimization time. Models were considered converged using the same criteria specified in the previous section, with $n=100$ simulation replicates.

\subsection{Phylogenetic trait imputation}
While models such as occupancy and N-mixture models are widely used to address ecological and population dynamics problems, we shift our focus in this section to mixed-type hierarchical models for evolutionary dynamics and introduce a novel approach to phylogenetic trait imputation (PTI).  We start with a dated phylogenetic tree, where we seek to estimate a vector of traits $\bm{z}_n =(z_{n1},z_{n2})$ for each vertex $n$, where $z_1$ is a discrete-valued trait and $z_2$ is a continuous-valued trait, $n\in \{1,...,n_{tip}\}$ indexes the tips, and $n\in \{ n_{tip}+1,...,N\}$ indexes the ancestral nodes.  The tree also includes $J$ edges $e_j$, where each edge has length $l_j$ and connects parent node $p[j] \in \{ n_{tip}+1,...,N\}$ and child node $c[j] \in \{ 1,2, ..., N \}$.   

We specifically define a state-switching process for the categorical trait and a Brownian motion model for the continuous trait.  To define this, we first introduce a ``one-hot encoding`` function $O(z_{n1})$, which converts the categorical trait with $K$ levels for node $n$ to a $K$ length indicator vector with a $1$ for the indicated level and $0$ elsewhere.  We then define a conditional distribution for the child node of edge $j$ given the value of their parent:

\begin{align}
    O(z_{c[j]1}) &\sim \mathrm{Categorical}\left( O(z_{p[j]1})^T e^{l_j\mat{Q}} \right) \\
    z_{c[j]2} &\sim \mathrm{Normal}\left( \vect{\beta} (O(z_{c[j]1}) - O(z_{p[j]1})) + z_{p[j]2} , l_j \sigma^2 \right)    
\end{align}

\noindent where $\mat{Q}$ is the $K \times K$ generator matrix for the instantaneous state-switching rate \citep{lewis_likelihood_2001}, and $\vect{\beta}$ is the impact of each level of the categorical trait on the continuous trait (where we fix $\beta_1=0$ for identifiability). Instantaneous transition rates $\mat{Q}$ is constructed such that rows sum to one, and $e^{l_j\mat{Q}}$ is the time-integrated transition probability over evolutionary time $l_j$ separating a parent and child node.  Scientific knowledge can then be used to structure $\mat{Q}$, e.g., where (1) a symmetric transition rate indicates an equal probability of each trait-transition, (2) a banded transition rate for $K \geq 3$ indicates that traits are naturally ordered such that level $K=1$ must transition to $K=2$ before moving to $K=3$, and (3) a blockwise sparse $\mat{Q}$ arises when modelling two or more categorical traits jointly.  We then complete the model by specifying a distribution for measurement errors:

\begin{align}
    O(y_{c[j]1}) &\sim \mathrm{Categorical}\left( O(z_{p[j]1})^T \mat{C} \right) \\
    z_{c[j]2} &\sim \mathrm{Normal}\left( z_{p[j]2}, \sigma_M^2 \right)    
\end{align}
where $\vect{y_n}$ is the vector of trait measurements, $\mat{C}$ is the probability of misclassifying a given trait (``confusion matrix``), and $\sigma_M$ is the standard deviation for measurement errors in the continuous-valued trait. This state-space formulation allows us to impute missing trait-values for ancestral nodes or tips, and also accomodates a known or estimated magnitude for measurement errors in both trait types.  Although other packages can also impute missing categorical traits \citep{gendre_benchmarking_2023}, our implementation requires less custom-built computational machinery than other recent developments \citep{lau_efficient_2026}.  

In the following, we demonstrate this mixed-type PTI model using previously published data for \textit{Liolaemus} lizards \citep{dominguez-guerrero_viviparity_2024}, using an ultrametric molecular phylogeny \citep{esquerre_how_2019}.  We extract viviparity $\vect{y}_1$ and female body size $\vect{y}_2$, and estimate a direct effect $\beta_2$ of viviparity on body size as well as one transition-rate parameter in the symmetric $2\times2$ rate matrix $\mat{Q}$ while assuming a trivially small misclassification probability and estimating the measurement-error parameter $\sigma_M$.  We then evaluate the strength and significance of the direct effect, and visually inspect the reconstructed traits for ancestral taxa.  

\section{Results}

\begin{figure}[htb]
    \centering
    \includegraphics[width=1\textwidth]{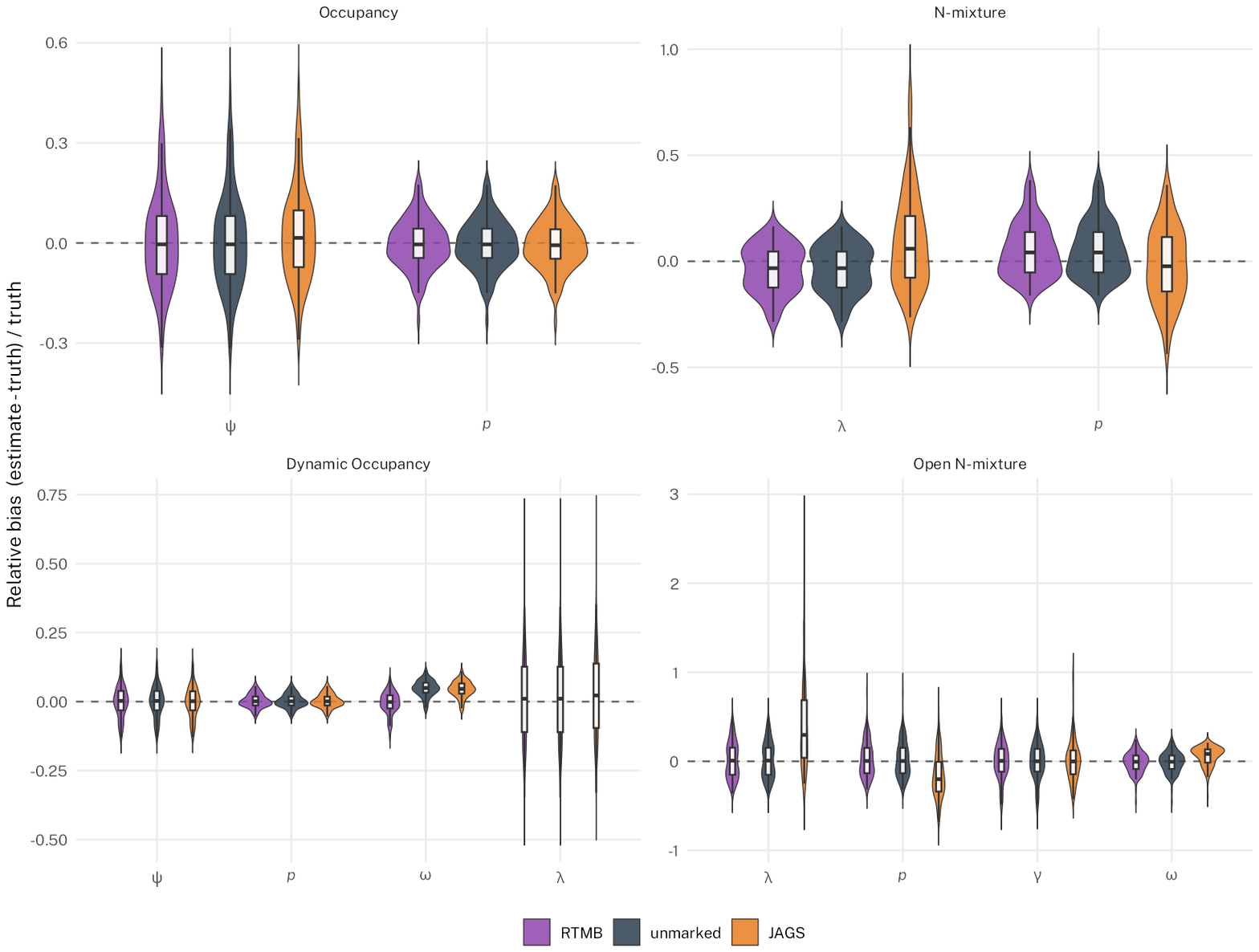}
    \caption{Distribution of relative error for parameter estimates (y-axis) for estimated parameters (x-axis) for the occupancy model (top-left panel), dynamic occupancy model (bottom-left), N-mixture (top-right), or open N-mixture (bottom-right).  See parameter definitions in the section \ref{sec:performance_comparison}
    }
    \label{fig:estimate_recovery}
\end{figure}

\begin{figure}[htb]
    \centering
    \includegraphics[width=1\textwidth]{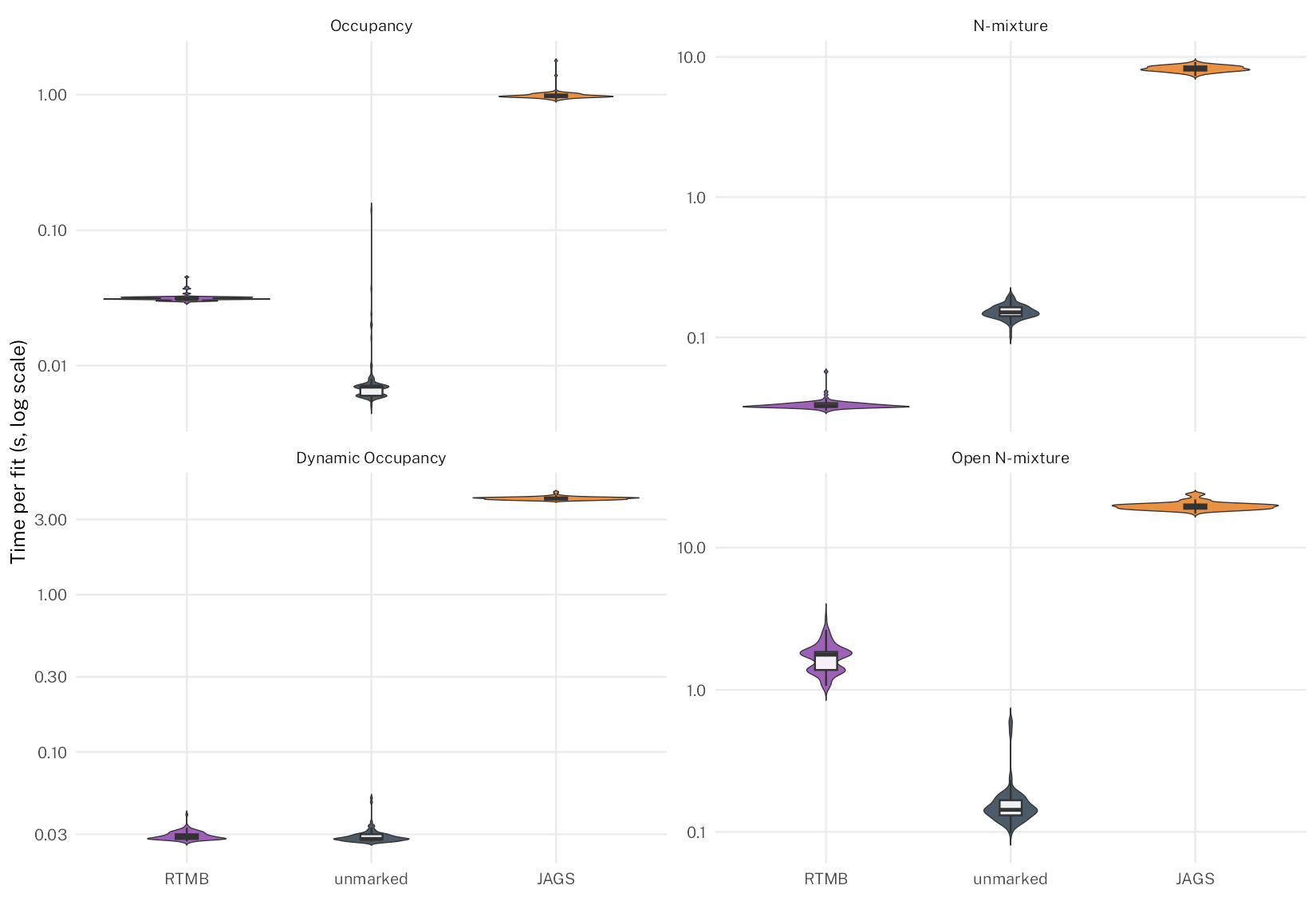}
    \caption{Run times (y-axis on log-scale) for each estimation model (x-axis) for the occupancy model (top-left panel), dynamic occupancy model (bottom-left), N-mixture (top-right), or open N-mixture (bottom-right)
    }
    \label{fig:timing_violin}
\end{figure}

Comparing parameter estimates using sequential reduction (using RTMB), single-use benchmark software (\texttt{unmarked}), and bespoke Gibbs sampling (using \texttt{JAGS}) confirms that all implementations give similar estimates (Fig. \ref{fig:estimate_recovery}).  The main exception is for the open N-mixture model (Fig. \ref{fig:estimate_recovery}, right panel), where Gibbs sampling results in much wide estimation intervals for initial abundance and recruitment, presumably resulting from the influence of priors combined with somewhat uninformative data about those parameters in some simulation replicates.  Despite \texttt{unmarked}'s use of custom implementations tailored to these optimization problems, the general-purpose RTMB implementation achieves comparable runtimes, with RTMB being faster for dynamic occupancy and N-mixture models and only slightly slower for occupancy and open N-mixture models (Fig. \ref{fig:timing_violin}).  This difference arises despite \texttt{unmarked} using custom-build code that is optimized for these estimation problems.  Similarly, MCMC sampling using JAGS (green violin in Fig. \ref{fig:timing_violin}) is 10-100 times slower than either maximum-likelihood estimator.  For these widely used demographic models, there appears to be little cost to using automated software for Sequential Reduction, while still allowing user-specified choices via purpose-built code.         

\begin{figure}[htb]
    \centering
    \includegraphics[width=1\textwidth]{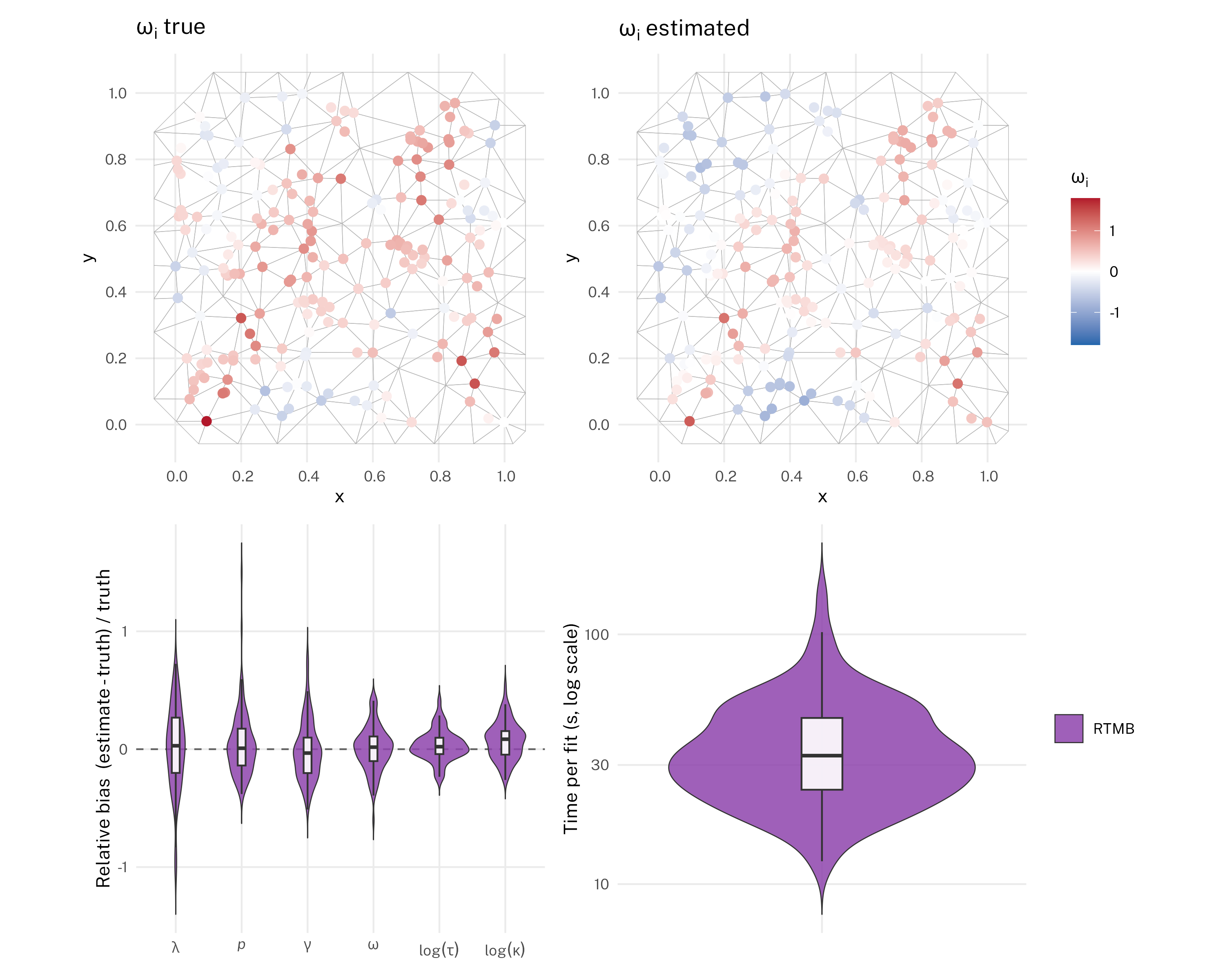}
    \caption{Simulated (top-left) and estimated (top-right) recruitment rate $log(\gamma_s) = \mu+\omega_s$ where $\vect{\omega}$ follows a Gaussian Markov random field, showing a single simulation replicate of a spatial open N-mixture model for dynamics at each site $s \in {1,2,...,S}$ (points) and also showing the SPDE mesh used to construct the precision matrix (grey triangles).  We also show the relative error for each model parameter (bottom-left), and the runtime (in seconds on log-scale; bottom-right) for estimating parameters and standard errors   
    }
    \label{fig:spde_field}
\end{figure}

Next, we visualize a single replicate from the novel spatial open N-mixture model, which estimates a spatial latent variable across $M=200$ sites via the Laplace approximation jointly with sequential reduction for the convolution of survival and recruitment over $T=4$ sampling intervals (Fig. \ref{fig:spde_field}, top row).  Inspecting results for a single replicate shows that site-specific recruitment can be accurately identified from an open N-mixture model, with higher recruitment in the western and eastern areas of the simulated domain.  Replicating this experiment, we confirm that the model runs in 10-100 seconds, and that the parameters are estimated centered around their true values (Fig. \ref{fig:spde_field}, bottom row).  We therefore conclude that the discrete-convolution for demographics can be combined with marginalizing across a continuous-valued (spatial) latent variable.  

\begin{figure}[htb]
    \centering
    \includegraphics[width=1\textwidth]{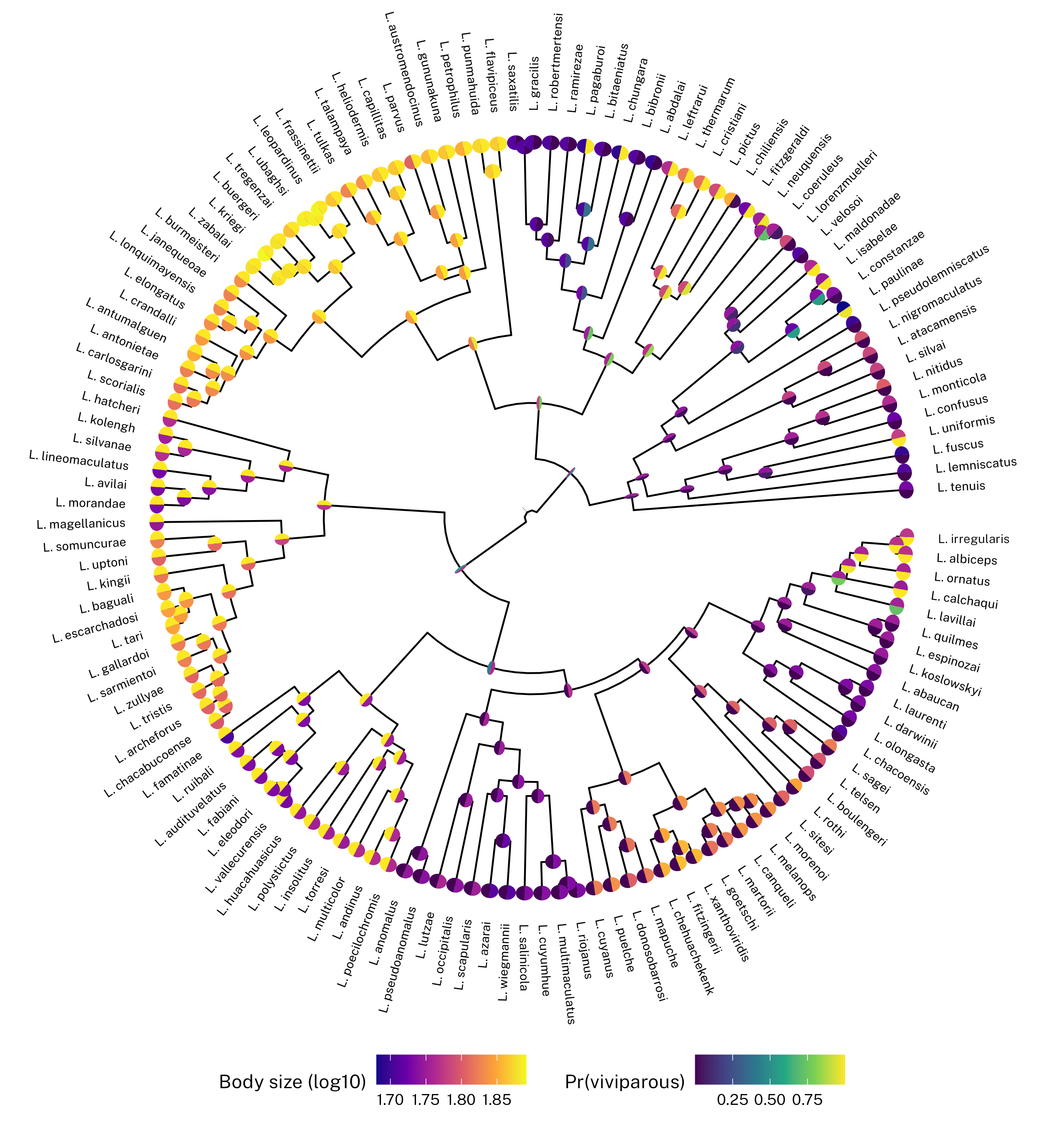}
    \caption{Radial phylogenetic tree (showing common ancestor in the middle, species as tips, and unobserved ancestors as interior nodes) for \textit{Liolaemus} lizards, including estimated body size and viviparity for both observed species and unobserved ancestors, showing the probability of viviparity (bounded from 0 to 1) on the right-hand-side (facing outward) for every taxon, as well as the log10 of body size on the left-hand side (facing outward) of each colored circle.   
    }
    \label{fig:phylogeny}
\end{figure}

Finally, we also visualize a novel phylogenetic trait imputation for a binary trait (viviparity) and its impact on a continuous trait (log10 body size) for a clade of lizards (Fig. \ref{fig:phylogeny}).  We specifically see two monophyletic groups (Fig. \ref{fig:phylogeny}, left-hand side) that are viviparous (i.e., yellow half-circle on right-hand-side facing outwards), and also generally have larger body sizes.  By contrast, other clades have a mix of viviparous and non-viviparous reproduction, and often have relatively smaller body sizes.  Importantly, the model also imputes these same traits for unobserved ancestral nodes (interior nodes in Fig. \ref{fig:phylogeny}), which provide a Markov bridge from the estimated root to each observed tip.  This combination of mixed-type traits and ancestral reconstruction ran in seconds, and is therefore feasible for small to medium-sized phylogenies such as this one.  

\section{Discussion}

We developed a new implementation for sequential reduction that can be efficiently combined with the Laplace approximation to fit ecological and evolutionary models that include a combination of discrete- and continuous-valued variables.  The implementation in RTMB automatically detects an efficient order to eliminate variables for the discrete convolution of demographic processes (e.g., representing birth/death processes), and then uses automatic differentiation to estimate continuous-valued parameters.  This automated process results in optimization that is comparable or faster than custom-built benchmark software (\texttt{unmarked}), and 10-100 times faster than Markov chain Monte Carlo sampling. High-level model specification also allowed us to develop two novel models that have not been previously published, which include complicated linkages among continuous variables (GMRFs for spatial and evolutionary correlations) and discrete-valued demographic or evolutionary dynamics. Despite these complex model structures, both novel models were specified in approximately 20–30 lines of user-specified R code that is then interpreted by RTMB, highlighting the compactness with which complex model structures can be expressed using this framework.

As with any new method, we see many avenues for further development and testing.  For example, we have presented the posterior mean estimates for discrete-valued variables (e.g., Fig. \ref{fig:phylogeny}) but RTMB also allows the posterior mode (i.e., the Viterbi algorithm) and future research could explore which of these two options to use when calculating model residuals.  Similarly, recent research measures the effective degrees of freedom for continuous-valued random effects e.g. for use in calculating the model parsimony for conditional predictions \citep{zheng_note_2024}, but it is unclear how to measure the parsimony of conditional predictions given discrete-valued variables without access to posterior samples \citep{watanabe_widely_2013}.  Finally, previous research has published alternative algorithms for identifying the optimal order for sequential reduction (see \citeauthor{hojsgaard_graphical_2012} \citeyear{hojsgaard_graphical_2012}) and future research could explore how different ordering methods compare for different dependency structures and model sizes.  Despite these technical details, the current implementation was fast and unbiased for medium-sized mixed-type models (e.g., 100-1000 random effects) that included complicated dependencies among variables, so we propose that the current implementation is likely practical for a wide range of applied and theoretical applications.  

We envision that easier access to mixed-type hierarchical models will allow bespoke development of a wide range of ecological and evolutionary models.  For example, Hidden Markov models using the forward-backwards algorithm are widely used for state-swiching in time, e.g., in phylogenetic or animal movement models \citep{mcclintock_multistate_2024}, and the Ising or Potts models are a natural generalization of these models to two or more dimensions \citep{noble_emergent_2015}, e.g., where ecological communities switch between alternative states across a two-dimensional landscape or specified site-connectivity matrix \citep{hanski_metapopulation_1998}.  Fitting the Ising model requires marginalizing across the set of potential states (i.e., the parameter-dependent normalization), which SR does automatically as a latent variable.  The latent assignment of sites to communities could then be fitted as a generalized linear mixed model to community sampling data, e.g., as a joint species distribution model with discrete community transitions.  Finally, animal movement models typically solve state-switching for each animal independently, which ensures that it can be solved as with one-dimensional dependencies.  However, recent research on coupled hidden Markov models allows state-switching to be associated with habitat patches or occur synchronously across animals \citep{pohle_primer_2021}.  In these cases, state-switching movement models could be extended to include multidimensional dependencies among animals, while automatically detecting an efficient order to eliminate variables. These examples represent only a small fraction of the potential applications of the sequential reduction approach implemented here. By making models with complex interactions between discrete and continuous processes easier to specify and fit, we hope this framework will facilitate the development of a broad range of purpose-built ecological and evolutionary models that would otherwise require considerable effort to develop and implement.

\bibliography{thorson_citations}

\clearpage

\section{Figures}

\begin{figure}[htb]
    \centering
    \includegraphics[width=1\textwidth]{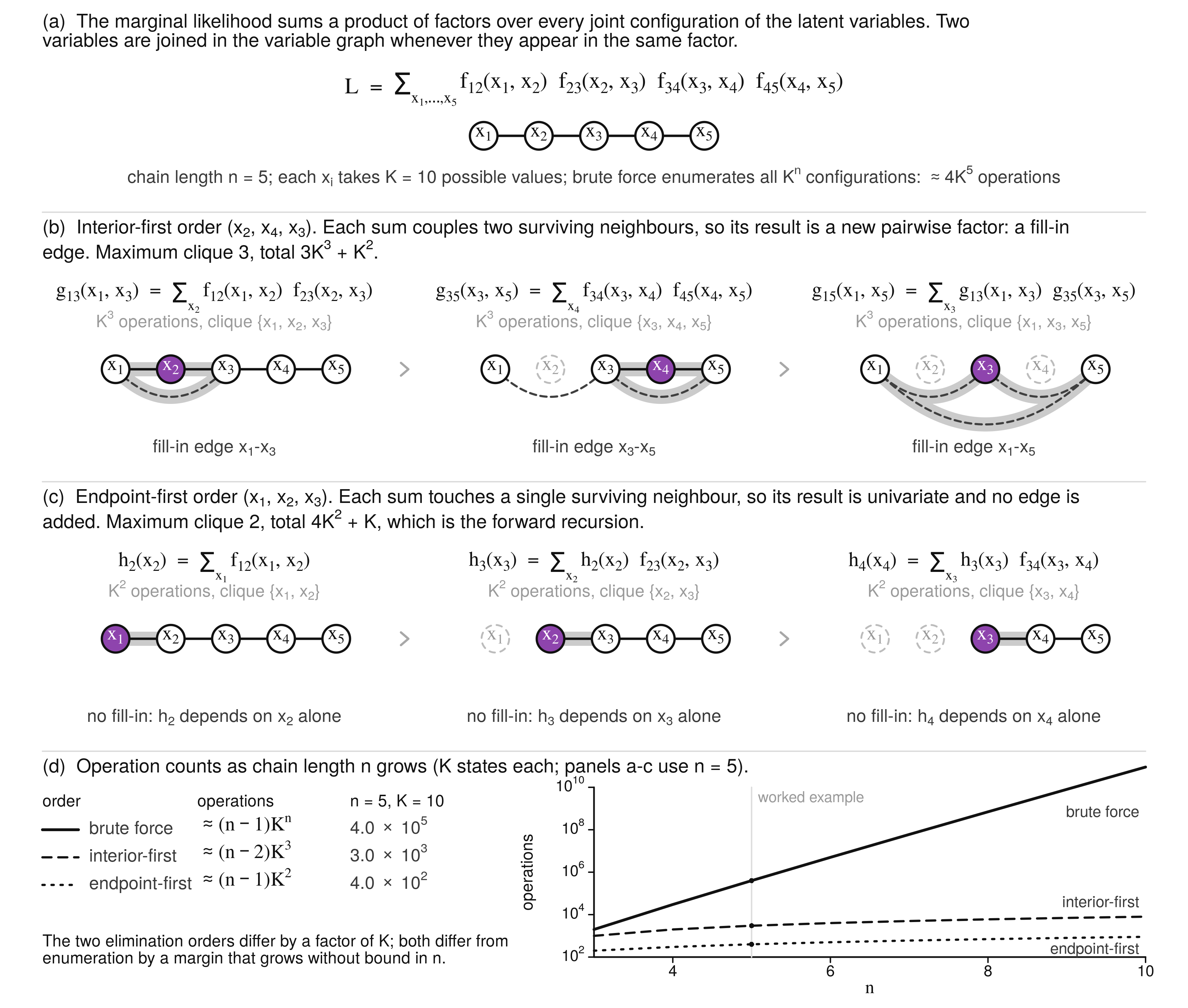}
    \caption{Sequential reduction as vertex elimination on the graph of a five-variable chain $x_1,\ldots,x_5$. (a) The marginal likelihood sums over every joint configuration of the latent variables. (b) A naive order that eliminates interior variables first creates fill-in edges and requires three-variable cliques, which is shown for contrast and is not the order RTMB uses. (c) For this chain, eliminating from an endpoint avoids fill-in entirely, using only two-variable cliques -- the forward recursion of a hidden Markov model. (d) Operation counts versus chain length $n$ ($K=10$ states per variable): both elimination orders scale polynomially in $n$, versus the exponential cost of brute-force enumeration. RTMB automates this process by finding and using an order like (c), although fill-in cannot always be avoided for more densely connected graphs.}
    \ref{fig:sequential_reduction}
    \label{fig:sequential_reduction}
\end{figure}

\clearpage

\clearpage

\clearpage

\clearpage

\clearpage

\renewcommand{\thefigure}{S\arabic{figure}}
\renewcommand{\thetable}{S\arabic{table}}
\renewcommand{\theequation}{S\arabic{equation}}
\setcounter{figure}{0}
\setcounter{table}{0}
\addtocontents{toc}{\protect\setcounter{tocdepth}{0}}

\onehalfspacing
\setcounter{page}{1}
\setcounter{equation}{0}

\begin{Center}
\section{Supporting Information S1}\label{sec:SI1}
\end{Center}

\noindent

The sequential reduction subsection showed that eliminating a discrete variable can induce new connections among its remaining neighbors (``fill-in''), increasing clique sizes and computational cost. Importantly, the amount of fill-in encountered by RTMB can depend on how random effects are grouped in the model specification. Thus, two parameterizations of the same statistical model may have substantially different computational costs: one may allow RTMB to identify an efficient elimination order with little fill-in, whereas another may constrain it to an order that creates larger intermediate cliques. We use the open N-mixture model to illustrate this distinction and to develop practical guidance for structuring random effects to avoid unnecessary fill-in.

Consider the open N-mixture model of Dail and Madsen (2011), with a state
variable $z_{jt}$ representing abundance at site $j$ and time $t$. The
model specifies
\[
z_{j1} \sim \mathrm{Poisson}(\lambda), \qquad
s_{jt} \sim \mathrm{Binomial}(z_{jt},\omega), \qquad
g_{jt} \sim \mathrm{Poisson}(\gamma),
\]
with abundance dynamics
\[
z_{j,t+1} = s_{jt} + g_{jt},
\]
and observations
\[
y_{jt} \sim \mathrm{Binomial}(z_{jt},p).
\]
Here, $\lambda$, $\gamma$, and $\omega$ are the initial abundance,
recruitment, and survival parameters, respectively, and $p$ is the
detection probability.

For sequential reduction (SR), consider the conditional-independence graph of
the discrete state variables. The factors involving survival and the
state transition connect $z_{jt}$ to $s_{jt}$ and $s_{jt}$ to
$z_{j,t+1}$, respectively. Thus, for each site, the conditional-independence graph has the form
\[
z_{j1} \;-\; s_{j1} \;-\; z_{j2} \;-\; s_{j2} \;-\; \cdots
\;-\; s_{j,T-1} \;-\; z_{jT}.
\]
The absence of edges between non-adjacent variables follows from the
conditional-independence structure of the model where successive abundance
states are connected through the corresponding survival variable and no
direct factor connects non-adjacent states.

The $z_{jt}$ and $s_{jt}$ are discrete random effects and therefore must be
passed to RTMB's SR machinery through the \texttt{integrate()} argument.
There are two ways to specify these random effects, but they are not computationally
equivalent. If survivors and abundances are declared as one combined random
effect vector, RTMB can search over the entire graph when choosing an elimination
order. It can therefore eliminate from an endpoint: $z_{j1}$ has a single
neighbor, $s_{j1}$, and each subsequent variable can be eliminated with
only one surviving neighbor. No fill-in is created, so the largest clique
has size two (panel c, Fig.~\ref{fig:sequential_reduction}), and computational performance 
is preserved.

If instead survivors and abundances are declared as two separate random
effect vectors -- the block of every $s_{jt}$ (\texttt{S}) and the block of every
$z_{jt}$ (\texttt{N}) -- \texttt{integrate()} eliminates each named entry as
a complete block. Thus, the \texttt{S} block is eliminated before the
\texttt{N} block. Each $s_{jt}$ then has two remaining neighbors, $z_{jt}$
and $z_{j,t+1}$, so eliminating $s_{jt}$ creates a fill-in edge between
them. The next survivor elimination therefore encounters a clique of size
three (panel b, Fig.~\ref{fig:sequential_reduction}). The endpoint-first
order is unavailable because $z_{j1}$ belongs to the separate \texttt{N}
entry and cannot be eliminated before the \texttt{S} entry. Below we demonstrate 
what these parameterizations look like in RTMB code for this specific model, 
and conduct a simple demonstration illustrating their computational costs. Readers 
unfamiliar with RTMB are encouraged to consult the introductory vignette:
\url{https://kaskr.r-universe.dev/RTMB/doc/RTMB-introduction.html}.

\subsection{Joint parameterization: survivors and abundances declared in one vector}

\begin{verbatim}
## data
J <- nrow(y)
T <- ncol(y)
K <- max(y) * 2
dat <- list(y = y, J = J, T = T)

## parameter list, with S and N combined into a single discrete random effect
par <- list(
  log_lambda = log(mean(y[, 1]) + 0.1),      ## starting values
  log_gamma = log(1.5),
  logit_omega = 0,
  logit_p = 0,
  SN = matrix(K, nrow = J, ncol = 2 * T - 1) ## LOOK HERE - good
)

# negative log-likelihood
f <- function(par) {
  getAll(par, dat)                           
  S <- SN[, seq_len(T - 1)]                  ## LOOK HERE - good
  N <- SN[, T - 1 + seq_len(T)]              ## LOOK HERE - good
  lambda <- exp(log_lambda)
  gamma <- exp(log_gamma)
  omega <- plogis(logit_omega)
  p <- plogis(logit_p)
  jnll <- -sum(dbinom(S, N[, 1:(T - 1)], omega, log = TRUE))
  jnll <- jnll - sum(dpois(N[, 1], lambda, log = TRUE))
  for (t in 1:(T - 1)) {
    G <- N[, t + 1] - S[, t]
    jnll <- jnll - sum(dpois(G, gamma, log = TRUE))
  }
  jnll <- jnll - sum(dbinom(y, size = N, prob = p, log = TRUE))
  jnll
}

## construct the AD function; S and N are integrated out jointly
obj <- MakeADFun(f, par,
  random = c("SN"),                    ## LOOK HERE - good
  integrate = list(
    SN = TMB::SR(0:K, discrete = TRUE) ## LOOK HERE - good 
  )
)

## optimize the marginal likelihood
opt <- nlminb(obj$par, obj$fn, obj$gr)
\end{verbatim}

\subsection{ Na\"{i}ve parameterization: survivors and abundances declared separately}

\begin{verbatim}
## data
J <- nrow(y)
T <- ncol(y)
K <- max(y) * 2
dat <- list(y = y, J = J, T = T)

## parameters, with S and N as separate discrete random effects
par <- list(
  log_lambda = log(mean(y[, 1]) + 0.1),
  log_gamma = log(1.5),
  logit_omega = 0,
  logit_p = 0,
  S = matrix(K, nrow = J, ncol = T - 1), ## LOOK HERE -- Bad
  N = matrix(K, nrow = J, ncol = T)      ## LOOK HERE -- Bad
)

## negative log-likelihood
f <- function(par) {
  getAll(par, dat)
  lambda <- exp(log_lambda)
  gamma <- exp(log_gamma)
  omega <- plogis(logit_omega)
  p <- plogis(logit_p)
  jnll <- -sum(dbinom(S, N[, 1:(T - 1)], omega, log = TRUE))
  jnll <- jnll - sum(dpois(N[, 1], lambda, log = TRUE))
  for (t in 1:(T - 1)) {
    G <- N[, t + 1] - S[, t]
    jnll <- jnll - sum(dpois(G, gamma, log = TRUE))
  }
  jnll <- jnll - sum(dbinom(y, size = N, prob = p, log = TRUE))
  jnll
}

## construct the AD function; S and N are integrated out separately
obj <- MakeADFun(f, par,
  random = c("S", "N"),                ## LOOK HERE - Bad
  integrate = list(
    S = TMB::SR(0:K, discrete = TRUE), ## LOOK HERE - Bad
    N = TMB::SR(0:K, discrete = TRUE)  ## LOOK HERE - Bad
  )
)

## optimize the marginal likelihood
opt <- nlminb(obj$par, obj$fn, obj$gr)
\end{verbatim}

\subsection{Comparing joint and separate parameterizations}

\begin{figure}[htb]
    \centering
    \includegraphics[width=1\textwidth]{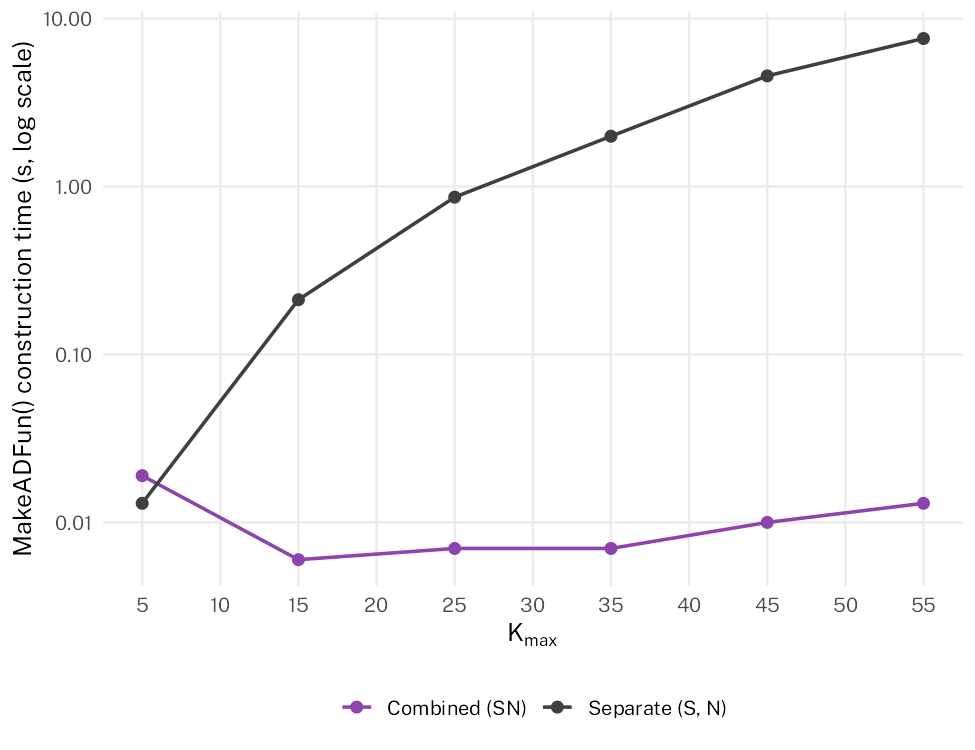}
    \caption{\texttt{MakeADFun()} construction time (y-axis, log scale) against truncation bound $K_{max}$ (x-axis) for the open N-mixture likelihood at $T=12$, comparing the combined ($SN$) and separate ($S,N$) discrete random-effect parameterizations.  The separate declaration's cost grows sharply with $K_{max}$ while the combined parameterization stays nearly flat. Code to reconstruct this figure is available at \texttt{R/elimination\_order\_timing.R} in the GitHub repository listed in the Data Availability Statement.
    }
    \label{fig:elimination_order_timing}
\end{figure}

We compare the difference between these parameterizations using one simulated realization of the open N-mixture model ($T=12$), timing only the construction of RTMB's \texttt{MakeADFun()}. This isolates the cost associated with the elimination order from the cost of subsequent optimization (Fig. \ref{fig:elimination_order_timing}). Consistent with the clique sizes identified above, construction time for the joint random-effects declaration remains approximately flat as the truncation bound $K$ increases, whereas construction time for the separate declaration grows sharply, consistent with the $O(K^2)$ versus $O(K^3)$ per-step costs implied by maximum clique sizes of two and three, respectively.

The general intuition is that RTMB's \texttt{integrate()} searches for an efficient elimination order \emph{within} a named entry rather than across entries. Consequently, discrete variables that appear together in a common factor of the joint likelihood should be declared as a single random effect. Doing so allows RTMB's elimination-order search to account for the full extent of their mutual dependence and can avoid unnecessarily large intermediate cliques.

\FloatBarrier
\clearpage

\end{document}